\documentclass{article}

\usepackage{PRIMEarxiv}

\usepackage[utf8]{inputenc} 
\usepackage[T1]{fontenc}    
\usepackage{hyperref}       
\usepackage{url}            
\usepackage{booktabs}       
\usepackage{amsfonts}       
\usepackage{nicefrac}       
\usepackage{microtype}      
\usepackage{lipsum}
\usepackage{tabularx}
\usepackage{fancyhdr}       
\usepackage{graphicx}       
\graphicspath{{media/}}     
\usepackage{algorithm}%
\usepackage{algorithmicx}%
\usepackage{algpseudocode}%
\usepackage{multirow}%
\usepackage{amsmath}
\title{APT-MCL: An Adaptive APT Detection System Based on Multi-View Collaborative Provenance Graph Learning
}

\author{
  Mingqi Lv \\
  College of Geoinformatics \\
  Zhejiang University of Technology\\
  Huzhou\\
  China\\
  \texttt{\{mingqilv@zjut.edu.cn} \\
   \And
  Shanshan Zhang \\
  College of Computer Science and Technology \\
  Zhejiang University of Technology \\
  Hangzhou\\
  China\\
  \texttt{211123120041@zjut.edu.cn} \\
     \And
  Haiwen Liu \\
  College of Computer Science and Technology \\
  Zhejiang University of Technology \\
  Hangzhou\\
  China\\
  \texttt{2112112262@zjut.edu.cn} \\
     \And
  Tieming Chen \\
  College of Computer Science and Technology \\
  Zhejiang University of Technology \\
  Huzhou\\
  China\\
  \texttt{tmchen@zjut.edu.cn} \\
     \And
  Tiantian Zhu \\
  College of Geoinformatics \\
  Zhejiang University of Technology \\
  Hangzhou\\
  China\\
  \texttt{ttzhu@zjut.edu.cn} \\
}

\begin{document}
\maketitle

\begin{abstract}
Advanced persistent threats (APTs) are stealthy and multi-stage, making single-point defenses (e.g., malware- or traffic-based detectors) ill-suited to capture long-range and cross-entity attack semantics. Provenance-graph analysis has become a prominent approach for APT detection. However, its practical deployment is hampered by (i) the scarcity of APT samples, (ii) the cost and difficulty of fine-grained APT sample labeling, and (iii) the diversity of attack tactics and techniques. Aiming at these problems, this paper proposes APT-MCL, an intelligent APT detection system based on Multi-view Collaborative provenance graph Learning. It adopts an unsupervised learning strategy to discover APT attacks at the node level via anomaly detection. After that, it creates multiple anomaly detection sub-models based on multi-view features and integrates them within a collaborative learning framework to adapt to diverse attack scenarios. Extensive experiments on three real-world APT datasets validate the approach: (i) multi-view features improve cross-scenario generalization, and (ii) co-training substantially boosts node-level detection under label scarcity, enabling practical deployment on diverse attack scenarios.
\end{abstract}


\section{Introduction}
APT (Advanced Persistent Threat) attacks are cyber-attacks initiated by well-organized attackers targeting critical industrial systems (e.g., energy, transportation, communication) and essential infrastructures (e.g., military, finance, health care). APT attacks can cause significant security threats, such as the exfiltration of sensitive data and compromise of system integrity. Therefore, the detection of APTs has become an important research topic in both academia and industry \cite{bib1,bib2}.
However, conventional “single point” cyber-attack detection strategies (e.g., malware detection \cite{bib3,bib4}, malicious traffic detection \cite{bib5,bib6}) are ineffective in detecting APT attacks. The reasons are twofold. First, APT attacks are persistent. Unlike traditional hit-and-run cyber-attacks, APT adversaries usually lurk within the target’s hosts and carry out attacks in a staged and covert manner. Second, APT attacks are stealthy. Attackers often exploit zero-day vulnerabilities and use a variety of attack tactics / techniques to evade the detection system. In response to these challenges, recent studies propose to leverage provenance data for APT detection \cite{bib7,bib8,bib9}. Provenance data, collected from the hosts using OS auditing frameworks, are typically represented as a directed acyclic graph (called a provenance graph), where the nodes stand for system entities (e.g., processes, files, sockets) and the edges denote system events (e.g., fork, write, open). The provenance graph provides a structured representation of contextual and causal relationships among system entities, which enables the effective tracing of the long-term and implicit behaviors of APT attacks.
Existing provenance-based APT detection can be divided into two categories: rule-based methods and the learning-based methods. Rule-based methods try to detect malicious behaviors using well-crafted rules \cite{bib10,bib11,bib12}. While these methods are efficient and easy to deploy, they suffer from the generalization and scalability problems (e.g., it is unable to detect unknown or mutant attacks). On the other hand, deep learning-based methods have the potential of discovering latent and fine-grained APT patterns and generalizing to unknown APT attacks. Although many previous studies on deep learning-based APT detection have achieved promising results \cite{bib13,bib14,bib15}, their practical deployment remains challenging due to the following obstacles.

\textbf{The scarcity of APT attack samples (C1):}APT attacks are rare events. It is almost impossible for a single organization to collect a large number of real APT attack samples, and different organizations would not share APT attack samples due to the consideration of privacy and sensitiveness. In addition, since APT attacks are highly complex, it is also extremely difficult to collect simulated APT attack samples on a large scale. For example, the most well-known simulated APT attack dataset, DARPA TC \cite{bib16}, also contains only a small number of APT attacks.

\textbf{The difficulty of sample labeling (C2):}A malicious behavior of APT attacks is usually concealed among tens of thousands of system entities in a provenance graph, and thus searching and labeling each malicious node in the provenance graph is like looking for a needle in a haystack. Furthermore, attackers may compromise hosts utilizing covert techniques such as process injection \cite{bib17} and file contamination \cite{bib18}, which further obscure the behavioral boundaries between malicious and benign system entities. These issues make it difficult to accurately label APT training samples in a fine-grained manner.

\textbf{The variety of APT attack strategies (C3):}Attackers employ a wide range of attack tactics and techniques to perform attack campaigns in different scenarios. Existing APT detection models, however, exhibit a notable deficiency in cross-domain adaptability. To demonstrate this issue, we conducted an experiment using ThreaTrace \cite{bib19}, a well-known anomaly detection model, on three different APT datasets, i.e., DARPA TC, StreamSpot \cite{bib20}, and our proprietary ransomware dataset, RWD. The models were trained and tested on the same datasets, yielding F1-scores of 0.965, 0.948, and 0.242, respectively. The reason of the poor performance on RWD is that ThreaTrace is primarily designed for general APT attacks in public datasets (e.g., DARPA TC and StreamSpot), while RWD is a private dataset containing ransomware attacks characterized by tactics and techniques that deviate significantly from those in DARPA TC and StreamSpot.
Aiming at these challenges, this paper proposes APT-MCL, an intelligent APT detection system based on Multi-view Collaborative provenance graph Learning. For challenges C1 and C2, APT-MCL adopts an unsupervised learning strategy to learn the normal system behavioral patterns from only benign samples, with no need of the labeled APT attack samples, and discovers APT attacks at the node level through anomaly detection (i.e., identifying system entities whose behavior greatly deviates from normal system behavioral patterns as anomalies). For challenge C3, APT-MCL creates multiple anomaly detection sub-models based on multi-view features and integrates them within a collaborative learning framework to identify diverse APT patterns. Importantly, these sub-models are independently initialized for each feature view, as we argue that directly concatenating all multi-view features into a single model risks overfitting, especially in an unsupervised learning setting. In summary, the main contributions of this paper are as follows.

  1) We propose an intelligent APT detection system based on unsupervised learning. It can learn normal system behavioral patterns from benign provenance graphs and detect node-level APT attacks in an anomaly detection way.
  
  2) We design multi-view features to cover multiple attack scenarios with different attack tactics and techniques. To ensure the universality of the multi-view features, each feature view is compatible and adaptable to most APT attack datasets.
  
  3) We propose a collaborative learning framework to integrate multiple sub-models trained on multi-view features. This framework progressively transitions the unsupervised learning sub-models into a supervised learning model by leveraging the complementary knowledge shared among these sub-models, thereby enhancing the model’s detection accuracy and generalization capability.
  
  4) We conducted extensive experiments on three real-world APT attack datasets. The experiment results show that multi-view features effectively capture richer behavioral information, and adapt to identify diverse attack scenarios. Moreover, the collaborative training framework achieves better and more stable detection performance across varying attack scenarios.

\section{Related work}
\label{{sec2}}
Existing provenance-based APT detection methods, based on their implementation details, can be categorized into two main groups: rule-based methods and learning-based methods.
\subsection{Rule-based APT Detection}
Rule-based approaches \cite{bib21,bib22,bib23} attempt to detect malicious behavior using manual-crafted rules. For example, SLEUTH \cite{bib24} identifies system entities and events that are most likely to be involved in APT attacks by using a label-based approach, first designing labels to encode the trustworthiness and sensitivity of subject and data, and then using these labels to design cyber-attack detection rules. Poirot detects threats by correlating a collection of metrics found by other systems and relies on expert knowledge of existing cyber threat reports to construct the attack graphs. It detects threats based on graph matching of provenance and attack graphs. However, it is difficult for them to detect unknown threats that are not included in  cyber threat reports. HOLMES \cite{bib25} is a layered framework for APT attack detection, where the key component is an intermediate layer that maps the underlying audit data to suspicious behaviors based on the rules of domain knowledge (e.g., ATT\&CK model). These rule-based detection methods offer notable advantages, such as high accuracy, efficiency, interpretability, and ease of deployment. Nonetheless, the design of such rules heavily relies on in-depth domain knowledge, which in turn make them prone to becoming outdated and exhibiting poor scalability when confronted with evolving APT attack strategies.
\subsection{Learning-based APT Detection}
\subsubsection{Supervised Learning Methods}
Another line of research \cite{bib26,bib27,bib28} utilizes learning-based strategies, including machine learning and deep learning techniques, to build APT attack detection models in an automated and intelligent manner. These methods exhibit the potential to uncover latent and fine-grained APT attack patterns, while also demonstrating the ability to generalize effectively to emerging APT attacks. For example, Barre et al. \cite{bib29} extracted a set of features from the provenance graph, such as the total amount of data written, the number of system files used, etc. to build a classifier to detect APT attacks. Xiang et al.\cite{bib30} extracted different features from both PC and mobile platforms, and then used multiple machine learning algorithms to detect APT attacks based on the combination of the features. Alsaheel et al.\cite{bib31} proposed ATLAS, an attack detection method based on the sequence representation of attack events within a graph. ATLAS combines the advantages of causal correlation analysis, natural language processing and machine learning to model both attack and benign behaviors. APT-KGL \cite{bib32} proposes an APT detection model based on heterogeneous graph learning, where the provenance data is represented as a heterogeneous graph. The model embeds nodes guided by a series of predefined meta-paths and distinguishes between malicious and benign process nodes through subgraph sampling and a classification model built using RGCN. Li et al.  \cite{bib33} proposed hierarchical APT detection method based on graph neural network with attention mechanisms, which identifies malicious events from intra-host provenance graphs within individual hosts and subsequently constructs inter-host provenance graphs at the network layer. By correlating host-level and network-level anomalies, the method effectively maps these anomalies to their corresponding APT stages. The above supervised learning-based methods have achieved superior detection results. However, their training processes are highly dependent on a large amount of annotated attack data, which entails the labor-intensive task of meticulously sifting through massive log data to label very few anomalous behaviors. Therefore, our work focuses on unsupervised APT detection methods.
\subsubsection{Unsupervised Learning Methods}
Unsupervised APT detection methods \cite{bib34,bib35,bib36} aim at identifying APT attack activities by analyzing network traffic, log data and system behaviors, etc., without the need of pre-labeled training data, which is commonly used for anomaly detection. Berrada et al.\cite{bib37} extracted Boolean-valued features from provenance graphs and regarded APT attack detection as an anomaly detection task through unsupervised learning techniques. StreamSpot \cite{bib38} conducts intrusion detection by analyzing information flow graph. It locally abstracts features of the graph to learn benign models and uses clustering methods to detect anomalous graphs. Unicorn \cite{bib39} further proposes a wl-kernel-based approach to extract features of the entire graph and learn evolutionary models to detect anomalous graphs. While it outperforms StreamSpot by utilizing contextualized graph analysis and evolutionary modeling, the inherent limitations of graph kernel methods hinder its ability to effectively detect stealthy threats. ThreaTrace is the first work to formalize the problem of host-based threat detection as the problem of detecting and tracking node-level anomalies in a provenance graph. They improved GraphSAGE to learn the combined normal characteristics of different node types to enable node-level detection. The core idea of ThreaTrace is that nodes associated with malicious behavior always have different patterns, which can be captured by deep learning algorithms. The disadvantage of unsupervised APT detection methods lies in their high false negative rate, particularly when confronted with the diversity and complexity of APT attack activities. To address this limitation, we note that malicious entities frequently exhibit unique structural patterns or sensitive behavioral traces contingent upon the type of attack executed. In light of this observation, we propose a multi-view anomaly detection strategy designed to independently capture both structural and behavioral features.
\section{Preliminary}
\subsection{Provenance Graph}
A provenance graph is defined as $G = (V, E)$, where the node set $V$ represents all system entities and the edge set $E$ represents all system events. Each edge $e = (u, v, a)$ indicates that a system entity $u$ (called subject) performs action $a$ on another system entity $v$ (called object). Figure~\ref{fig1} illustrates the snapshots of two provenance graphs, where ovals represent processes, rectangles represent files and registries, and the diamonds represent sockets.

\subsection{Threat Model}
In this paper, we focus on detecting APT-related malicious system entities within a host. Since processes are the primary initiators of APT attacks \cite{bib40,bib41,bib42}, our focus is on identifying malicious process nodes in a provenance graph. We assume that attackers could use various attack strategies (e.g., different tactics and techniques) to launch APT attacks, but they cannot compromise the OS kernel or the audit module. It means that the audit logs are trustworthy and any malicious activities can be captured by the audit module. In addition, we assume that adversaries would inevitably leave behavioral traces during the execution of attacks, and the behavioral patterns of malicious nodes would exhibit distinctive features that set them apart from benign nodes.
\subsection{Motivation}
\label{sec:33}
\begin{figure}[h]
\includegraphics[width=\textwidth]{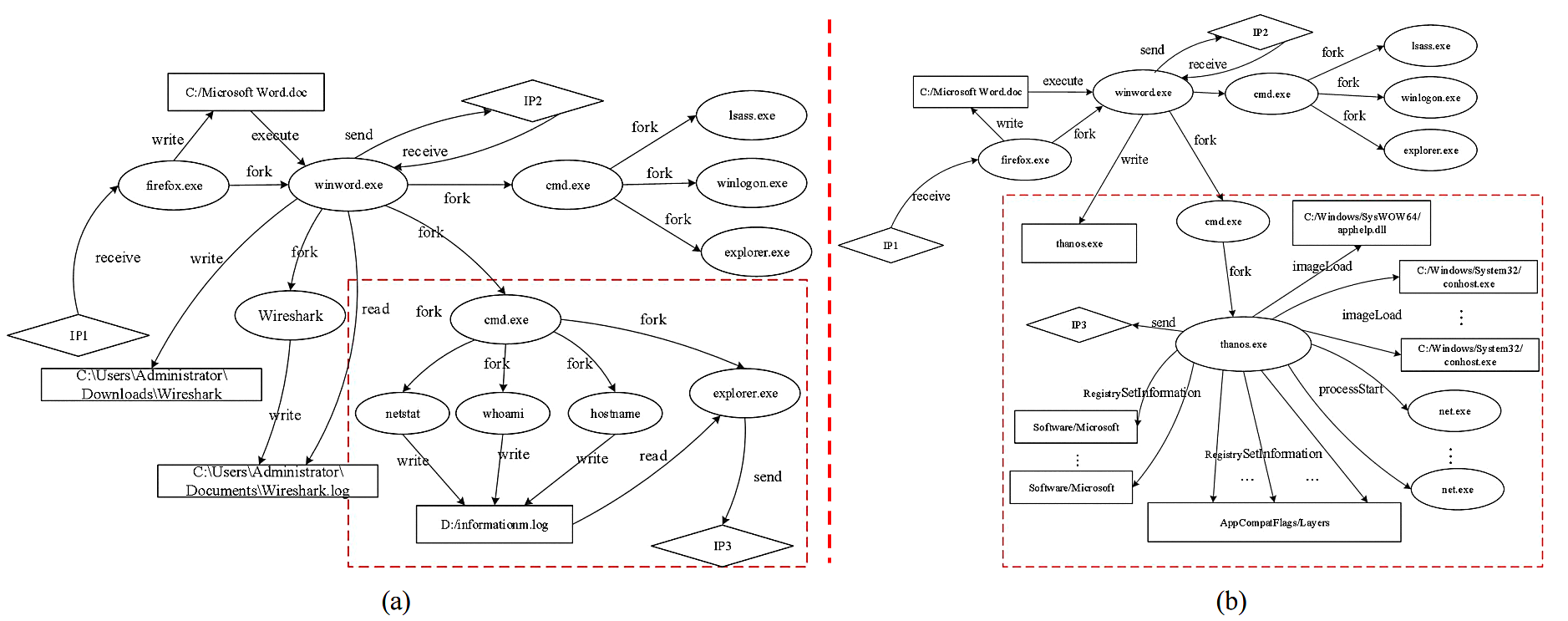}
\caption{Two examples of provenance graphs: (a) a provenance graph of a Collection\&Exfiltration attack; (b) a provenance graph of a ransomware attack.} \label{fig1}
\end{figure}

We present two cases to demonstrate the diversity of APT attack strategies, which motivate our work. The first case shown in Figure ~\ref{fig1}(a), represents a provenance graph of a Collection \& Exfiltration attack \cite{bib43}. The subgraph enclosed within the dotted rectangle highlights the malicious nodes associated with this attack. First, the malware program, disguised as a Word document (i.e., “winword.exe”), invokes “cmd.exe” to gather sensitive data from the local system. These collected data are subsequently dumped into a file “information.log” and transmitted to a remote C\&C server. Compared to benign system entities, the core malicious system entity “winword.exe” exhibits a distinctive characteristic, i.e., it engages in multiple sensitive activities, such as executing sensitive instructions (e.g., whoami, netstat, etc.), collecting sensitive data, and accessing a remote server.
The second case in Figure ~\ref{fig1}(b) represents a provenance graph of a ransomware attack. In this scenario, the malware program “winword.exe” executes a ransomware program “thanos.exe”. Then, “thanos.exe” conducts extensive I/O operations, including loading images, reading and modifying registry keys, and reading and encrypting files. By further analyzing the provenance data of this case, we found that the ransomware program executes tens of thousands of registry operations and file operations, which serves as a key distinguishing characteristic that sets it apart from benign system entities.

Obviously, the malicious system entities in the two cases demonstrate significantly different characteristics. If we were to use the features of sensitive activities (such as employed by APT-KG) to train the detection model, it would fail to detect the malicious system entities in the second case, as they hardly perform any sensitive activities. On the other hand, if we used features of graph structure (such as adopted by ThreaTrace) to train the detection model, it would struggle to effectively detect the malicious system entities in the first case, as these entities do not show a significant deviation in the number of specific types of edges within the provenance graph. This highlights that certain types of features are not universally adaptable to different attack scenarios. Worse still, since our goal is to train the detection model in an unsupervised manner, simply concatenating different types of features could divert the attention of the detection model from effective feature subsets and thus increases the risk of overfitting.
\subsection{System Architecture}
\begin{figure}[h]
\includegraphics[width=\textwidth]{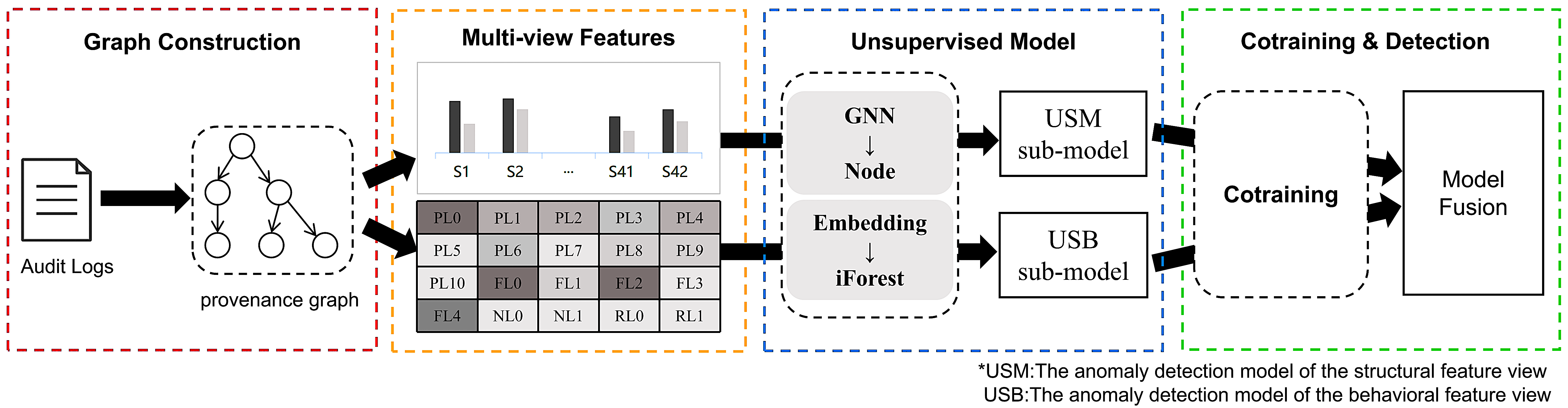}
\caption{Overview of APT-MCL.} \label{fig2}
\end{figure}

The architecture of APT-MCL is outlined in Figure ~\ref{fig2}, consisting of four modules.

\textbf{Provenance Graph Construction Module}(Section~\ref{sec:41}): It continuously collects system events through system auditing framework, and then extracted essential interactions to construct a provenance graph.

\textbf{Multi-view Feature Extraction Module}(Section~\ref{sec:42}): In order to accommodate different attack scenarios (C3), it extracts features for each system entity from multiple views. Specifically, we consider two feature views (i.e., the structural features and the behavioral features).

\textbf{Unsupervised Detection Module}(Section~\ref{sec:43}): Without relying on APT samples (C1), it trains anomaly detection sub-models for each view using only benign samples in an unsupervised manner. Additionally, to adapt to the stealthy and persistent characteristics of APT attacks, it adopts a GNN (Graph Neural Network) to extract higher level features that capture the causal and contextual correlations between distant graph neighbors.

\textbf{Collaborative Learning Module}(Section~\ref{sec:44}): Unsupervised learning models are prone to overfitting when faced with diverse attack scenarios (C2). To address this challenge, a customized collaborative learning framework is designed to generate pseudo labels, thereby converting the unsupervised learning task into a semi-supervised one. This module combines all sub-models to produce more accurate detection results.
\section{Methodology}
\subsection{Provenance Graph Construction}
\label{sec:41}
We use KELLECT \cite{bib44}, an efficient provenance data collection tool, to collect kernel-level system events. Each collected system event is initially parsed as a triad (subject, action, object), and then modeled as two nodes connecting by an edge within the provenance graph. Since each system entity is assigned a unique identifier, the “object” of one system event and the “subject” of another system event that reference the same system entity can be merged as a single node. By doing so, the provenance graphs capture causal relationships between system entities and facilitate the reasoning over system entities that are temporally distant. As a result, provenance graphs prove invaluable for navigating and analyzing the stealthy and persistent APT attacks.
\subsection{Multi-View Feature Extraction}
\label{sec:42}
As discussed in Section~\ref{sec:33}, a single feature view struggles to effectively identify different attack scenarios. To tackle this limitation, we extract features for each system entity from multiple views to accommodate a wide range of attack scenarios. The design of multi-view feature follows three key principles. First, the feature views should be pervasive, meaning they can be computed for most provenance datasets. Second, each individual feature view should be independent, capable of supporting the training of the detection models on its own. Third, the feature views should be diverse, representing patterns  that correspond to different attack scenarios. By following these principles, we design two views of features, i.e., structural features and behavioral features.
\subsubsection{Structural features}According to the findings in \cite{bib45}, malicious nodes typically exhibit different interaction pattern with their neighboring nodes as compared with the benign nodes, resulting in distinct local graph structures. Specifically, the interaction pattern of a node can be represented as the distribution of various types of edges connected to it. For example, in a ransomware attack, numerous file operations are often generated, which is reflected in the provenance graph as a significant concentration of “process-to-file” edges in the local area.

Based on this insight, we extract structural features for each node as follows. First, we categorize the nodes in a provenance graph into four types (including process, file, registry, and socket), and the interactions between different node pairs can form 21 edge types (as shown in Table~\ref{tab1}). Then, for each node $v_i$ in the provenance graph, we create a 42-dimensional vector $\mathbf{s}_{e_i}=[a_1, a_2, \ldots, a_{21}, a_{22}, a_{23}, \ldots, a_{42}]$ as the structural features of $v_i$, where $a_i$ $(1 \le i \le 21)$ is the number of incoming edges of $v_i$ with the $i$-th type and $a_j$ $(22 \le j \le 42)$ is the number of outgoing edges of $v_i$ with the $(j-21)$-th type. For example, the structural feature vector of the node winword.exe in Figure~\ref{fig1}(b) is $
[1, 0, 0, 0, 0, 0, 0, 0, 0, 0, 0, 0, 1, 0, 0, 0, 0, 1, 0, 0, 0, 0, 0, 0, 1, 0, 0, 0, 0, 0, 0, 1, 0, 0, 0, 0, 0, 0, 1,\allowbreak 0, 0, 0]$.

\begin{table}[h]
\caption{The summary of monitored events.}\label{tab1}
\centering
\begin{tabular}{@{}ll@{}}
\toprule
\textbf{Node Type Pairs} & \textbf{Edge Types}                                                     \\ \midrule
process - process        & launch                                                                  \\ \midrule
process - file           & create, read, write, close, delete, open                                \\ \midrule
process - registry       & open, query, enumerate, modify, close, delete                           \\ \midrule
process - socket         & send, receive, retransmit, copy, connect, disconnect, accept, reconnect \\ 
\bottomrule 
\end{tabular}
\end{table}
\subsubsection{Behavioral features}Focusing solely on structural features can easily result in false positives, as both benign and malicious processes might display similar local structures. For example, a legitimate process that backs up files and a malicious process that performs data exfiltration could exhibit similar distribution of edge types, such as regularly reading files and sending data to a remote server. On the other hand, APT attacks are often characterized by performing sensitive behaviors, such as accessing sensitive data or invoking sensitive instructions. For example, during the Reconnaissance stage of an APT attack, sensitive instructions like whoami, systeminfo, and tasklist are often executed to collect information from a host.

Therefore, we extract behavioral features for each node as follows. First, we define 16 sensitive behavioral labels corresponding to different entity types, as detailed in Table~\ref{tab2} \cite{bib46}. Here, an IoC (Indicator of Compromise) is a piece of digital forensic evidence indicating that a host has been breached (e.g., an IP address, URL, file hash, etc.). Based on this, for each node $v_i$ in the provenance graph, we create a 20-dimensional vector $\mathbf{b}_{e_i} = [b_1, b_2, \ldots, b_{20}]$ as the behavioral features of $v_i$, where $b_i$ ($1 \le i \le 20$) is a Boolean indicator of the $i$-th label.

\begin{table}[h]
\caption{The summary of behavioral features.}\label{tab2}
\begin{tabularx}{\textwidth}{@{} lXX @{}}
\toprule
\textbf{Labels} & \textbf{Description}                      & \textbf{Node} \\ \midrule
PL0             & The system entity is a process            & process       \\ 
PL1             & Process has network connection            & process       \\ 
PL2             & Process reads files from network          & process       \\ 
PL3             & Process modifies files from network       & process       \\ 
PL4             & Process executes files from network       & process       \\ 
PL5             & Process interacts with non-existent files & process       \\ 
PL6             & Process reads sensitive files             & process       \\ 
PL7             & Process modifies sensitive files          & process       \\ 
PL8             & Process executes sensitive files          & process       \\ 
PL9             & Process executes sensitive instructions   & process       \\ 
PL10            & Process modifies registry items           & process       \\ \midrule
FL0             & The system entity is a file               & file          \\ 
FL1             & File is sensitive                         & file          \\ 
FL2             & File is uploaded                          & file          \\ 
FL3             & File contains data from network           & file          \\ 
FL4             & File contains IoC attributes              & file          \\ \midrule
NL0             & The system entity is a socket             & socket        \\ 
NL1             & Socket contains IoC attributes            & socket        \\ \midrule
RL0             & The system entity is a registry           & registry      \\ 
RL1             & Registry contains IoC attributes          & registry      \\ \bottomrule
\end{tabularx}
\end{table}
\subsection{Unsupervised Detection Model Building}
\label{sec:43}
We build the unsupervised anomaly detection model through two stages: the feature learning stage and the model training stage.

In the feature learning stage, the initial feature vectors of a node extracted in Section~\ref{sec:42} can only reflect its first-order structural/behavioral patterns. However, APT attacks are stealthy, which often conceals for a long period of time and launches the attacks by using alternative system entities. Therefore, the first-order structural/behavioral patterns fail to capture long-distance causal correlations, thus leading to reduced detection precision. To address this limitation, we employ a GNN to aggregate information from each node’s neighbors for the learning of higher-order structural/behavioral patterns. Given the absence of malicious/benign labels for nodes in this task, we train the GNN in a self-supervised way. In this approach, node types inherently present in the provenance graph are utilized as the supervised signals, and the GNN is trained on a downstream task of node type classification (i.e., classifying a node into process, file, registry, or socket).

Using its intrinsic message-passing mechanism, the GNN model can be trained by stacking multiple layers to capture distant interaction semantics. In the $t$-th layer, the node feature matrix $E^{(t)}$ is updated according to Equation~\ref{eq:1}, where $G$ is the topology of the provenance graph (expressed as an adjacency matrix), $W^{(t-1)}$ is a trainable parameter matrix, and $GNN(\cdot)$ serves as a GNN-based encoder that can be instantiated by various GNN models (e.g., GCN \cite{bib47}, GAT \cite{bib48}, GIN \cite{bib49}, etc.). Importantly, increasing the depth of the GNN allows it to capture higher-order structural and behavioral patterns.

\begin{equation}
\mathbf{E}^{(t)}=G N N\left(\mathbf{W}^{(t-1)}, G, \mathbf{E}^{(t-1)}\right)
\label{eq:1}
\end{equation}

For each individual feature view, we stack $T$ GNN layers on the provenance graph $G$ to learn the final node feature matrix $E^{(T)}$, which is then input into a softmax classifier to make node type classification based on Equation~\ref{eq:2}, where $P$ is the probability distribution of the classification and $W^{(T)}$ is the parameter matrix of the classifier. The training loss is calculated as the cross-entropy between $P$ and the actual node type distribution $Y$. The $GNN(\cdot)$ is then optimized by minimizing this loss. After training, $GNN(\cdot)$ can be utilized as a feature extractor, which takes the provenance graph $G$ and its initial node feature matrix $E^{(0)}$ as input, and produces the final node feature matrix $E^{(T)}$ as output.

\begin{equation}
\mathbf{P}=\operatorname{softmax}\left(\mathbf{E}^{(T)} \cdot \mathbf{W}^{(T)}\right)
\label{eq:2}
\end{equation}
During the model training stage, we apply iForest \cite{bib50}, an effective outlier detection algorithm, to construct individual anomaly detection model for each feature view. Each anomaly detection model takes the final feature vectors of all the “process” nodes as input and outputs anomaly scores for each “process” nodes. Nodes with higher anomaly scores (i.e., outliers) are identified as potential malicious nodes.
\subsection{Collaborative Detection Model Training}
\label{sec:44}
\subsubsection{Collaborative training}
Upon constructing the anomaly detection model of the structural feature view (abbreviated as $USM$) and the model of the behavioral feature view (abbreviated as $UBM$), we fuse them through our customized collaborative training framework to derive the final detection model. The objective of multi-view knowledge fusion is to leverage complementary information from multiple perspectives to enhance the model's robustness and generalization capabilities across diverse attack scenarios. Broadly, there are two mainstream strategies for achieving multi-view knowledge fusion: model-level fusion and feature-level fusion \cite{bib51}. The model-level fusion strategy comprehensively integrates the outputs of all sub-models to make the final decision, while the feature-level fusion strategy consolidates all multi-view features into a unified feature vector for model detection. Unfortunately, both two strategies face significant limitations in our task. First, the two sub-models (i.e., $USM$ and $UBM$) are trained in unsupervised way (i.e., the malicious/benign labels are unavailable). As a result, most model-level fusion methods (e.g., Boosting, Stacking, Dense Layer, etc.) cannot be applied, since these methods commonly rely on calculating classification probabilities across different labels or evaluating the classification accuracies on labeled subsets. While a simple majority voting scheme, such as Bagging \cite{bib52}, could be employed for output combination, it treats all sub-models equally and thus fails to identify which sub-model is better suited for the current attack scenario. Second, due to the unsupervised nature of the training process, the feature-level fusion methods cannot determine which features are effective for a given system entity by learning the correlations between features and labels. Consequently, these methods are prone to be influenced by ineffective features, which can lead to overfitting and degrade the model's performance.

Based on the above analysis, the core challenge that hinders the multi-view knowledge fusion for our task is that the unsupervised nature of the sub-model training. Aiming at this issue, we try to convert the unsupervised learning task into a supervised learning style by exploiting the complementary knowledge of the two feature views to collaboratively provide supervised signals to each other. Specifically, we design a collaborative training framework based on Co-Training algorithm \cite{bib53}. Co-Training is a semi-supervised learning algorithm, which assumes that each sample can be featured by two independent views. Ideally, the two views can provide complementary knowledge. Thus, if a sample is misclassified by one view, it might be correctly classified by the other view. Co-Training algorithm can leverage the complementary knowledge to collaboratively refine the sub-models.

As illustrated in Algorithm~\ref{alg:1}, our training framework firstly applies the anomaly detection models trained in Section~\ref{sec:43} to find a small batch of samples that could be mostly confidently identified as malicious and benign, and then trains two sub-models (i.e., $SM$ and $BM$) in a supervised way by treating the small batch as a labeled sample set. Second, it refines the two sub-models by assigning pseudo labels to the unlabeled samples and providing them to each other in a gradual and iterative way. Specifically, we define two parameters, i.e., $\mathit{batch\_thres}$ and $\mathit{batch\_prob}$. In each iteration, we treat samples whose classification confidence greater than $\mathit{batch\_thres}$ as high confidence samples, from which we then select a proportion of $\mathit{batch\_prob}$ to assign pseudo labels.
\begin{algorithm}[!htbp]
\caption{The collaborative training framework}
\label{alg:1}
\footnotesize
\begin{algorithmic}[1]

\Statex \textbf{Input:}
\Statex \hspace*{\algorithmicindent}\parbox[t]{0.93\linewidth}{The unlabeled provenance node sample dataset, $UD$.}
\Statex \hspace*{\algorithmicindent}\parbox[t]{0.93\linewidth}{The anomaly detection model of the structural feature view, $USM$.}
\Statex \hspace*{\algorithmicindent}\parbox[t]{0.93\linewidth}{The anomaly detection model of the behavioral feature view, $UBM$.}
\Statex \hspace*{\algorithmicindent}\parbox[t]{0.93\linewidth}{The classification confidence threshold, $\mathit{batch\_thres}$.}
\Statex \hspace*{\algorithmicindent}\parbox[t]{0.93\linewidth}{The sample batch probability, $\mathit{batch\_prob}$.}

\Statex \textbf{Output:}
\Statex \hspace*{\algorithmicindent}\parbox[t]{0.93\linewidth}{The supervised sub-model of the structural feature view, $SM$.}
\Statex \hspace*{\algorithmicindent}\parbox[t]{0.93\linewidth}{The supervised sub-model of the behavioral feature view, $BM$.}

\ForAll{process node $s_i \in UD$}
    \State \parbox[t]{0.93\linewidth}{Use $USM$ to detect $s_i$, and output the malicious/benign label and anomaly probability of $s_i$.}
    \State \parbox[t]{0.93\linewidth}{Use $UBM$ to detect $s_i$, and output the malicious/benign label and anomaly probability of $s_i$.}
\EndFor

\State \parbox[t]{0.93\linewidth}{$SMA \gets$ Get all samples in $UD$ that $USM$ outputs anomaly probability $> \mathit{batch\_thres}$.}
\State \parbox[t]{0.93\linewidth}{$BMA \gets$ Get all samples in $UD$ that $UBM$ outputs anomaly probability $> \mathit{batch\_thres}$.}

\State \parbox[t]{0.93\linewidth}{$SMS \gets$ Select $\mathit{batch\_prob}\times |SMA|$ samples in $SMA$ with the highest anomaly probabilities.}
\State \parbox[t]{0.93\linewidth}{$SBS \gets$ Select $\mathit{batch\_prob}\times |SMA|$ samples in $SMA$ with the lowest anomaly probabilities.}
\State \parbox[t]{0.93\linewidth}{$BMS \gets$ Select $\mathit{batch\_prob}\times |BMA|$ samples in $BMA$ with the highest anomaly probabilities.}
\State \parbox[t]{0.93\linewidth}{$BBS \gets$ Select $\mathit{batch\_prob}\times |BMA|$ samples in $BMA$ with the lowest anomaly probabilities.}

\State \parbox[t]{0.93\linewidth}{Remove $SMS$, $SBS$, $BMS$, and $BBS$ from $UD$.}
\State \parbox[t]{0.93\linewidth}{$LD \gets$ malicious set $SMS \cup$ benign set $SBS \cup$ malicious set $BMS \cup$ benign set $BBS$.}

\While{$UD$ is not empty}
    \State \parbox[t]{0.93\linewidth}{$SM \gets$ Train in a supervised way based on $LD$ with the structural feature view.}
    \State \parbox[t]{0.93\linewidth}{$BM \gets$ Train in a supervised way based on $LD$ with the behavioral feature view.}

    \State \parbox[t]{0.93\linewidth}{Use $SM$ and $BM$ to classify each sample in $UD$.}

    \State \parbox[t]{0.93\linewidth}{$SMA \gets$ Get all samples in $UD$ that $SM$ outputs anomaly probability $> \mathit{batch\_thres}$.}
    \State \parbox[t]{0.93\linewidth}{$BMA \gets$ Get all samples in $UD$ that $BM$ outputs anomaly probability $> \mathit{batch\_thres}$.}

    \State \parbox[t]{0.93\linewidth}{$SMS \gets$ Select $\mathit{batch\_prob}\times |SMA|$ samples in $SMA$ with the highest anomaly probabilities.}
    \State \parbox[t]{0.93\linewidth}{$SBS \gets$ Select $\mathit{batch\_prob}\times |SMA|$ samples in $SMA$ with the lowest anomaly probabilities.}
    \State \parbox[t]{0.93\linewidth}{$BMS \gets$ Select $\mathit{batch\_prob}\times |BMA|$ samples in $BMA$ with the highest anomaly probabilities.}
    \State \parbox[t]{0.93\linewidth}{$BBS \gets$ Select $\mathit{batch\_prob}\times |BMA|$ samples in $BMA$ with the lowest anomaly probabilities.}

    \State \parbox[t]{0.93\linewidth}{Remove $SMS$, $SBS$, $BMS$, and $BBS$ from $UD$.}
    \State \parbox[t]{0.93\linewidth}{$LD \gets$ malicious set $SMS \cup$ benign set $SBS \cup$ malicious set $BMS \cup$ benign set $BBS$.}
\EndWhile

\end{algorithmic}
\end{algorithm}
\subsubsection{Sub-model fusion}
After the collaborative training, we obtain two sub-models that can independently classify process nodes based on the two feature views, respectively. However, due to their variations in discriminative power across different attack scenarios, a mechanism is required to effectively fuse the outputs of these two sub-models for detection. For example, the sub-model of the structural feature view usually has stronger discriminative power on APT attacks that exhibit significant deviation in system event distributions, whereas the sub-model of the behavioral feature view usually has stronger discriminative power on APT attacks that perform certain sensitive behaviors. To address this, we define the following four strategies for sub-model fusion, using process node $v_0$ as an example:

\textbf{Fusion Strategy 1} (BV, Benign Voting): The final classification result of $v_0$ will be malicious, if both sub-models classify $v_0$ as malicious.

\textbf{Fusion Strategy 2} (MV, Malicious Voting): The final classification result of $v_0$ will be malicious, as long as one sub-model classifies $v_0$ as malicious.

\textbf{Fusion Strategy 3} (SV, Soft Voting): It considers the classification probabilities provided by each sub-model. Each sub-model produces a probability distribution for classifying $v_0$ as either malicious or benign, resulting in two probability vectors: $[\mathit{MP}_{SSM}, \mathit{BP}_{SSM}]$ and $[\mathit{MP}_{SBM}, \mathit{BP}_{SBM}]$. Here, $\mathit{MP}_{SSM}$ and $\mathit{MP}_{SBM}$ denote the probabilities of sub-models $SSM$ and $SBM$ classifying $v_0$ as malicious, respectively. Similarly, $\mathit{BP}_{SSM}$ and $\mathit{BP}_{SBM}$ represent the probability of sub-models $SSM$ and $SBM$ classifying $v_0$ as benign. Then, we calculate the average probabilities: $\mathit{MP} = (\mathit{MP}_{SSM} + \mathit{MP}_{SBM})/2$ and $\mathit{BP} = (\mathit{BP}_{SSM} + \mathit{BP}_{SBM})/2$. If $\mathit{MP} > \mathit{BP}$, the $v_0$ is classified as malicious. Otherwise, it is classified as benign.

\textbf{Fusion Strategy 4} (ST, Stacking): We train a meta-model that takes the probability vectors (i.e., $[\mathit{MP}_{SSM}, \mathit{BP}_{SSM}]$ and $[\mathit{MP}_{SBM}, \mathit{BP}_{SBM}]$) generated by the two sub-models as input and outputs the final classification result. Since the labeled samples are not available, we create a pseudo labeled sample set, where the two sub-models have identical classification result.

\section{Experiment}
\subsection{Experiment Setup}
\subsubsection{Dataset}
We use the following three datasets that cover different APT scenarios to evaluate APT-MCL.

\textbf{DataBreach}: We simulated three APT attack scenarios on a Linux host through a red team campaign. The first, “attack by webshell”, exploits web vulnerabilities to penetrate the target system using webshell scripts. The second, “attack by RAT (Remote Access Trojan)”, implants Trojans via phishing attacks to steal sensitive information. and the third, “attack by LotL (Living off the Land)”, leverages legitimate software to execute malicious shellcode directly in memory.

\textbf{Ransomware}: It contains solely ransomware attacks targeting a Window host, and we simulated two types: “attack by phishing” and “attack by vulnerability”. The first infiltrates the system via email attachments containing executable malicious files, and then implants ransomware (e.g., Thanos, Cerber, Clop, etc.) into the target system. The second exploits system vulnerabilities to perform privilege escalation and execute ransomware payloads.

\textbf{DARPA TC E3}: It is the most prevalent public APT dataset \cite{bib54}, which was collected from DARPA’s Transparent Computing program. During the engagement, a red team launched a series of APT attacks towards an enterprise network, while engaging in benign activities such as browsing websites, checking emails, and SSH logins. This dataset covers multiple operation systems (i.e., Linux, Windows, and BSD) and various attack scenarios, including sensitive information exfiltration, malicious HTTP request, and malicious file download and execution. We choose three sub-datasets, including TRACE, THEIA, and CADETS, for our evaluation.
We summarized the detailed distribution of benign and malicious nodes for these datasets in Table ~\ref{tab3}. It is evident that each dataset contains varying amounts of attack entities, reflecting diverse APT strategies and scenarios.

\begin{table}[h]
\caption{Overview of  Datasets.}\label{tab3}
\begin{tabularx}{\textwidth}{@{} XXXX @{}}
\toprule
Dataset & Benign Nodes & Malicious Nodes & Edges \\
\midrule
DataBreach       & 29018                   & 714                        & 2700912          \\
Ransomware       & 21765                   & 159                        & 642346           \\
TRACE            & 2416007                 & 67383                      & 6978024          \\
THEIA            & 319446                  & 25319                      & 102929710        \\
CADETS           & 706966                  & 12852                      & 8663569          \\
\bottomrule 
\end{tabularx}
\end{table}
\subsubsection{Experiment Setting Up}
For the experiment environment, all experiments were implemented in Python using PyTorch on an Ubuntu 20.04 server with an Intel Xeon Gold 5218 CPU, 128 GB RAM, and an NVIDIA GeForce RTX 4090 GPU.

For the default parameter settings, the detection models adopt a two-layer GraphSAGE architecture, projecting input features into a 32-dimensional hidden space and then reducing to 15 dimensions, followed by a two-layer MLP classifier. ReLU activation and a dropout rate of 0.5 were used to balance accuracy and generalization. Models were trained using the Adam optimizer with a learning rate of 0.01 and weight decay of $5 \times 10^{-4}$ for 30 epochs per round, with a mini-batch size of 5{,}000 nodes and neighbor sampling to handle large graphs. The negative log-likelihood loss was employed.

For the testing strategy, the normal provenance nodes of each dataset are splitted into training set (denoted as $NTS$) and testing set (denoted as $NES$) according to a ratio of 7:3, and the malicious provenance nodes are also splitted into training set (denoted as $MTS$) and testing set (denoted as $MES$) according to the same ratio. Then, $NTS$ is used to train the sub-models of both views. $NTS$ and $MTS$ are used as the unlabeled sample dataset in the collaborative training phase. Finally, $NES$ and $MES$ are used to evaluate the models.

\subsection{Experiment 1: Parameter Tuning Experiment}
There are two key parameters in the collaborative detection model training module of APT-MCL, i.e., the classification confidence threshold to determine high-confidence sample batches, denoted as $\mathit{batch\_thres}$, and the probability of selecting unlabeled samples in each collaborative training iteration, denoted as $\mathit{batch\_prob}$. In this section, we investigate the impact of these two parameters using the DARPA TC dataset. We select ST as the fusion strategy.

\begin{figure}[h]
\includegraphics[width=\textwidth]{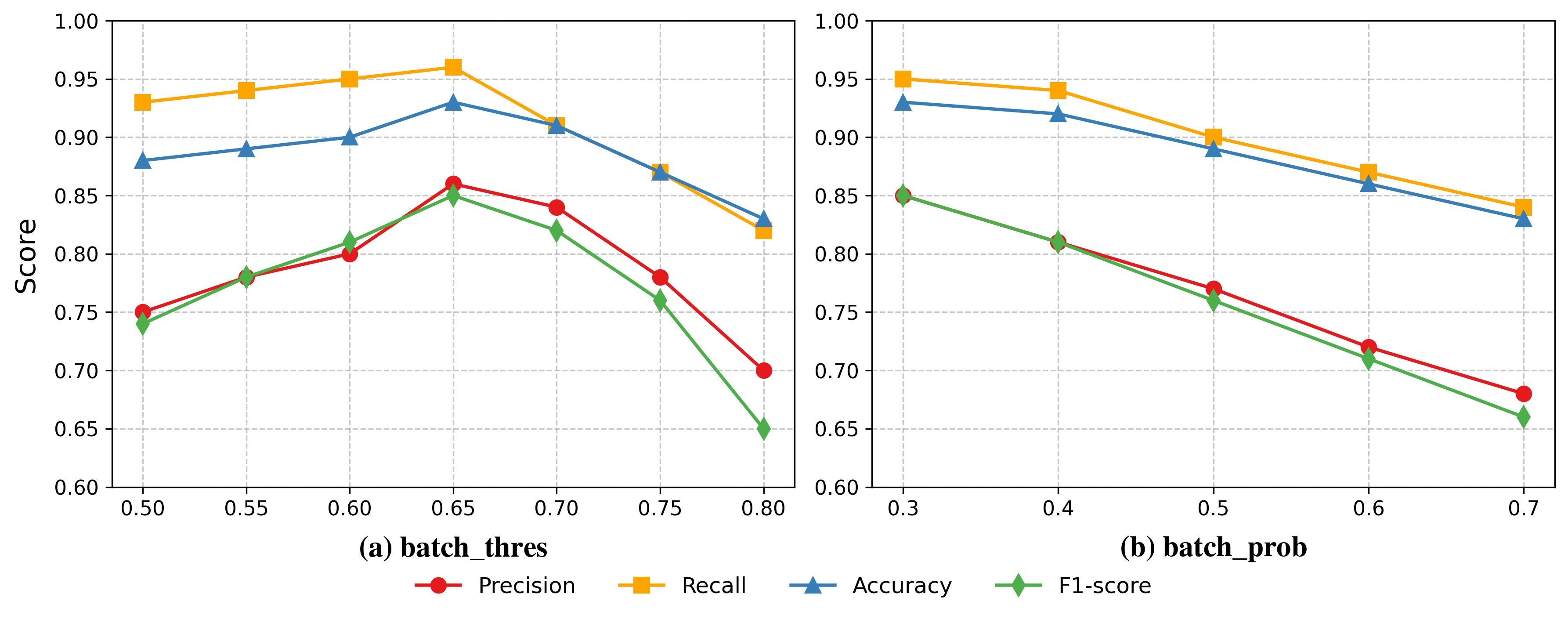}
\caption{Experiment results of parameter tuning: (a) tuning of parameter \texttt{batch\_thres}; (b) tuning of parameter \texttt{batch\_prob}.} \label{fig3}
\end{figure}

In the first experiment, we evaluate the effect of $\mathit{batch\_thres}$ ranging from 0.5 to 0.8 while fixing $\mathit{batch\_prob}=0.3$. The results are shown in Figure~\ref{fig3}(a). As $\mathit{batch\_thres}$ increases, the detection performance first improves and then drops significantly. When $\mathit{batch\_thres}$ is set too low, noisy samples may be introduced into sub-model training due to incorrect pseudo-labels in early iterations. When $\mathit{batch\_thres}$ is set too high, sub-models may fail to provide sufficient pseudo-labeled samples to sustain collaborative training. For example, with $\mathit{batch\_thres}=0.8$, the sub-model based on the behavioral-feature view provides no pseudo-labeled samples in the first iteration. The best overall detection performance is achieved at $\mathit{batch\_thres}=0.65$.

In the second experiment, we evaluate the effect of $\mathit{batch\_prob}$ ranging from 0.3 to 0.7. The results are shown in Figure~\ref{fig3}(b). Overall performance decreases steadily as $\mathit{batch\_prob}$ increases, particularly in Precision and F1-score. This is because the initial sub-models are weak, as they are trained in an unsupervised manner, and thus tend to generate incorrect pseudo-labels; consequently, a larger $\mathit{batch\_prob}$ introduces more noisy samples. Moreover, a larger $\mathit{batch\_prob}$ may also increase the risk of converging to a poor local optimum, whereas a smaller $\mathit{batch\_prob}$ increases the number of iterations required for convergence. Therefore, we set $\mathit{batch\_prob}=0.3$ to balance detection performance and computational overhead.
\subsection{Experiment 2: Ablation Experiment}
In the first experiment, we try to compare the performance of the four sub-model fusion strategies (BV, MV, SV, ST) in different APT attack scenarios, and explore how to improve the robustness of the detection system by fusing the classification results of the structural feature and behavioral feature sub-models. In this experiment, we use the DARPA TC E3 dataset to compare the performance of different integration strategies. Based on the experimental results presented in Table ~\ref{tab4}, we derive the following observations:

First, BV (Benign Voting) is the most conservative strategy. Although it can reduce false positives, it sacrifices the recall, resulting in relatively low F1-score. Second, although MV (Malicious Voting) guarantees almost perfect recall, it sacrifices the precision, also resulting in low F1-score. Third, SV (Soft Voting) takes into account the classification probability and thus is able to balance the output of the two sub-models, which achieves high precision and recall. It can provide better recall and F1-score, but the precision and accuracy are slightly inferior to ST (Stacking). Based on a detailed analysis, we found that most SV misclassifications fall into two categories. On one hand, false positives occur when benign processes display high-frequency of file or network operations (e.g., logging, monitoring daemons). In these situations, both behavioral and structural sub-models assign moderate anomaly scores, but the SV’s averaging mechanism amplifies this shared bias, incorrectly classifying them as malicious. In contrast, ST learns from the confidence patterns and inter-view disagreement to recognize such inconsistencies and correctly reclassify them as benign. On the other hand, false negatives appear in stealthy attack nodes, whose local behaviors seem benign but dependency structures link to malicious activities. SV tends to overlook these mixed-behavior nodes, since the benign behaviors dominate the averaged probability, while ST can dynamically adjust its weights through the meta-model, capturing the hidden structural cues and successfully identifying such latent threats.
\begin{table}[h]
\caption{The experiment results of sub-model fusion strategies.}\label{tab4}
\begin{tabularx}{\textwidth}{@{} l XXXX l @{}}
\toprule
Sub-model Fusion Strategy                                          & Precision & Recall & ACC & Macro-F1 \\ 
\midrule
\begin{tabular}{@{}c@{}}Fusion Strategy 1\\ (BV, Benign Voting)\end{tabular}    & 0.907                   & 0.738               & 0.805        & 0.795             \\ 
\midrule
\begin{tabular}{@{}c@{}}Fusion Strategy 2\\ (MV, Malicious Voting)\end{tabular} & 0.625                   & 0.999               & 0.981        & 0.769             \\
\midrule
\begin{tabular}{@{}c@{}}Fusion Strategy 3\\ (SV, Soft Voting)\end{tabular}      & 0.846                   & 0.921               & 0.864        & 0.823             \\ 
\midrule
\begin{tabular}{@{}c@{}}Fusion Strategy 4\\ (ST, Stacking)\end{tabular}         & 0.854                   & 0.953               & 0.904        & 0.847             \\ 
\bottomrule
\end{tabularx}
\end{table}

In the second experiment, we try to investigate the impact of various combinations of feature views as follows on the model’s detection performance.

(1) SFV: It refers to the detection model based only on the structural feature view.

(2) BFV: It refers to the detection model based only on the behavioral feature view.

(3) CON: It firstly concatenates the features from the structural view and behavioral view into unified single feature vectors, and then trains the detection model based on the unified feature vectors.

The experimental results are shown in Table ~\ref{tab5}. First, it can be seen that single-view features are insufficient for adapting to different attack scenarios. For example, while SFV slightly outperforms BFV on the Ransomware dataset, it has extremely poor performance on the DataBreach dataset, where the malicious activities are sparse and thus the edge type distributions of malicious nodes are not significantly different from those of benign nodes, limiting the structural view. In contrast, Ransomware induces file/registry I/O bursts that markedly shift these distributions, making structural features informative. Second, CON has only slight improvement over SFV and BFV, but still has significant performance gap compared to our method. It reveals that directly combining different views of features would distract the detection model from focusing on the most effective feature subset.
\begin{table}[h]
\caption{View features cross-scene experiment results.}\label{tab5}
\begin{tabularx}{\textwidth}{@{} l XXXX l @{}}
\toprule
Dataset & Variant & Precision & Recall & ACC & Macro-F1 \\
\midrule
\multirow{4}{*}{Ransomware} 
& SFV     & 0.4223 & 0.9719 & 0.9323 & 0.5888 \\
& BFV     & 0.3816 & 0.9987 & 0.9192 & 0.5524 \\
& CON     & 0.3654 & 0.9987 & 0.9134 & 0.5352 \\
& APT-MCL & 0.7403 & 0.5902 & 0.9955 & 0.6575 \\
\midrule
\multirow{4}{*}{DataBreach} 
& SFV     & 0.0383 & 0.0466 & 0.8146 & 0.0422 \\
& BFV     & 0.4178 & 0.5597 & 0.8812 & 0.4784 \\
& CON     & 0.4948 & 0.6100 & 0.9371 & 0.5464 \\
& APT-MCL & 0.7804 & 0.9005 & 0.9915 & 0.8327 \\
\bottomrule
\end{tabularx}
\end{table}

In the third experiment, we try to evaluate the performance of different training strategies. Specifically, we compare APT-MCL with the following two variants.

(1) Supervised: It takes the concatenated structural-behavioral feature vector as input, and trains a RandomForest classifier to learn the decision boundary between the benign and malicious nodes in a fully supervised way.

(2) Unsupervised: It also takes the concatenated structural-behavioral feature vector as input, and trains a iForest model solely on benign provenance nodes in a fully unsupervised way. Then, the model assigns an anomaly score to each node, flagging as suspicious those nodes that significantly deviate from the learned normal pattern.

Table ~\ref{tab6} summarizes the experiment results. First, a consistent ranking is observed, i.e., Supervised$>$APT-MCL$>$Unsupervised. This ranking reflects that the detection model could greatly benefit from the availability of fine-grained node-level labels via supervised learning. However, these fine-grained node-level labels are usually unavailable in practical usage. Second, comparing the three variants, we note that the supervised model exploits ground-truth node labels, while both the unsupervised variant and APT-MCL operate without manual annotations. The unsupervised iForest baseline trains a single detector on a concatenated structural-behavioral feature vector and relies purely on density-based anomaly scores, which tends to yield high recall at the cost of more false positives. In contrast, APT-MCL explicitly separates the structural and behavioral views into two sub-models and iteratively exchanges high-confidence pseudo labels between them. This cross-view agreement injects supervision-like signals and suppresses view-specific noise, leading to consistently higher precision and Macro-F1 while preserving high recall.

On the Ransomware dataset, where malicious behaviors are relatively salient, the supervised model achieves the best performance, and APT-MCL performs only slightly below the supervised counterpart while clearly outperforming the unsupervised baseline. On the DataBreach dataset, which comprises more complex and stealthy attacks, supervised learning again yields the strongest results, and APT-MCL remains highly competitive in the absence of annotations. Overall, across two distinct scenarios, APT-MCL substantially enhances detection compared with unsupervised learning and, although marginally inferior to full supervision, constitutes a practical and label-efficient solution for real-world APT detection. In particular, it avoids the need for expensive fine-grained annotations and can be more easily scaled and adapted to new environments where labeled data are scarce.
\begin{table}[h]
\caption{Experimental results of different learning strategies in Ransomware dataset and DataBreach dataset.}\label{tab6}
\begin{tabularx}{\textwidth}{@{} l XXXX l @{}}
\toprule
Dataset  & Model  & Precision & Recall & ACC  & Macro-F1 \\
\midrule
\multirow{3}{*}{Ransomware} & Unsupervised & 0.3654    & 0.9987 & 0.9134 & 0.5352   \\
                            & Supervised   & 0.9963    & 0.9987 & 0.9998 & 0.9949   \\
                            & APT-MCL      & 0.7948    & 0.9601 & 0.9911 & 0.8697   \\
\midrule
\multirow{3}{*}{DataBreach} & Unsupervised & 0.4948    & 0.6100 & 0.9371 & 0.5464   \\
                            & Supervised   & 0.9739    & 0.9069 & 0.8785 & 0.9355   \\
                            & APT-MCL      & 0.8176    & 0.6784 & 0.9619 & 0.7183  \\
\bottomrule
\end{tabularx}
\end{table}
\subsection{Experiment 3: Comparison Experiment}
In order to verify the advantages of APT-MCL, we compare it with the following three state-of-the-art baseline methods.

(1) Log2vec \cite{bib55}: It employs a heterogeneous graph embedding technique to capture the behavioral patterns, and then detects malicious system entities through clustering-based anomaly identification.

(2) MAGIC \cite{bib56}: It leverages masked graph representation learning to model nodes and edges of provenance graphs, and then identifies anomalous system entities via outlier detection methods.

(3) ThreaTrace : It utilizes GraphSAGE to perform self-supervised learning via a node classification task, and then detects anomalous system entities by considering the misclassified samples.

The experiment results are presented in Table ~\ref{tab7}. First, Log2vec performs reasonably on Ransomware, but its clustering boundary becomes unstable under weak signals on DataBreach, where malicious processes trigger few actions and their local structures are similar with those of benign nodes. Second, MAGIC, which learns representations via masked feature/structure reconstruction, is generally competitive on the DARPA subsets, where higher-order structural regularities are informative. However, its structure-centric single-model design underutilizes behavior-dominant bursts in Ransomware and struggles when malicious neighborhoods remain look-alike to benign ones in DataBreach. Third, ThreaTrace shows a consistent limitation across datasets, i.e., expanding misclassifications to two-hop neighborhoods inflates false positives most visibly under bursty benign activity in Ransomware, while reliance on local edge-type counts under-detects the subtle, near-benign malicious neighborhoods characteristic of DataBreach. It can be competitive in selected DARPA scenarios but remains prone to FPR inflation in dense or noisy regions. Fourth, APT-MCL achieves stable performance across all evaluated scenarios. Despite pronounced class imbalance within each train/test split, its results are comparable to—or better than—those of existing unsupervised APT detectors. By jointly exploiting structural and behavioral views, the multi-view representation captures complementary signals, increases the semantic separability of node embeddings, and thereby improves discrimination between malicious and benign entities.

Finally, our results indicate that the collaborative learning framework substantially improves the model’s capacity to learn malicious behaviors across different scenarios. When a given view starts weak, pseudo labels supplied by the other view(s) provide effective supervision and strengthen that sub-model’s detection. We further observe that, even when a sub-model initially yields very few pseudo labeled instances, the impact on retraining its peers is minimal, underscoring the framework’s inherent robustness.
\begin{table}[h]
\caption{Experimental results on the datasets.}\label{tab7}
\centering
\begin{tabularx}{\textwidth}{@{} l XXXX l @{}}
\toprule
Detector   & Dataset                     & Precision & Recall & FPR   & Macro-F1 \\
\midrule
Log2vec    & \multirow{4}{*}{Ransomware} & 0.418     & 0.527  & 0.999 & 0.422    \\
MAGIC      &                             & 0.792     & 0.999  & 0.981 & 0.771    \\
ThreaTrace &                             & 0.683     & 0.999  & 0.978 & 0.751    \\
APT-MCL    &                             & 0.828     & 0.674  & 0.013 & 0.743    \\
\midrule
Log2vec    & \multirow{4}{*}{DataBreach} & 0.294     & 0.672  & 0.117 & 0.319    \\
MAGIC      &                             & 0.637     & 0.972  & 0.056 & 0.714    \\
ThreaTrace &                             & 0.407     & 0.970  & 0.074 & 0.573    \\
APT-MCL    &                             & 0.845     & 0.964  & 0.005 & 0.900    \\
\midrule
Log2vec    & \multirow{4}{*}{CADETS}     & 0.490     & 0.850  & 0.016 & 0.621    \\
MAGIC      &                             & 0.940     & 0.990  & 0.002 & 0.970    \\
ThreaTrace &                             & 0.900     & 0.990  & 0.002 & 0.950    \\
APT-MCL    &                             & 0.999     & 0.999  & 0.000 & 0.999    \\
\midrule
Log2vec    & \multirow{4}{*}{THEIA}      & 0.622     & 0.661  & 0.003 & 0.642    \\
MAGIC      &                             & 0.980     & 0.990  & 0.001 & 0.990    \\
ThreaTrace &                             & 0.870     & 0.990  & 0.001 & 0.932    \\
APT-MCL    &                             & 0.864     & 0.973  & 0.012 & 0.915    \\
\midrule
Log2vec    & \multirow{4}{*}{TRACE}      & 0.541     & 0.780  & 0.018 & 0.640    \\
MAGIC      &                             & 0.990     & 0.990  & 0.001 & 0.990    \\
ThreaTrace &                             & 0.720     & 0.990  & 0.011 & 0.830    \\
APT-MCL    &                             & 0.999     & 0.996  & 0.000 & 0.998   \\
\bottomrule
\end{tabularx}
\end{table}
\subsection{Experiment 4: Computation Overhead Experiment}
We report the computational footprint of APT-MCL to contextualize deployment in Table ~\ref{tab8}. As a two-view architecture with iterative co-training, APT-MCL incurs more training and inference overhead than single-model baselines. On our hardware, the end-to-end cost remains moderate: provenance-graph construction and model updates fit within commodity GPU memory, and per-batch inference is suitable for offline triage workloads. In practice, parallelizing the view-specific encoders, caching per-entity features, and reusing precomputed graph statistics further amortize the cost. While APT-MCL is not optimized for speed, its runtime and memory footprint remain within the operational budget for host-based provenance analytics.

\begin{table}[h]
\caption{Memory utilization of APT-MCL.}\label{tab8}
  \centering
\begin{tabular}{@{}lll@{}}
\toprule
Method  & Avg. Training Time (s) & Peak Memory Consumption (MB)   \\
\midrule
APT-MCL & 2,470                  & \begin{tabular}[c]{@{}c@{}}Graph Construction: 1439\\ Training: 1590\\ Inference: 2084\end{tabular} \\
\midrule
MAGIC   & 1940                   & \begin{tabular}[c]{@{}c@{}}Graph Construction: 672\\ Training: 810\\ Inference:1532\end{tabular}   \\
\bottomrule
\end{tabular}
\end{table}

\section{Conclusions}
This paper introduces APT-MCL, an intelligent APT detection system based on unsupervised learning, which is applicable to various APT scenarios and demonstrates excellent detection performance. APT-MCL can learn normal system behavioral patterns from benign provenance graphs through multi-view features and collaborative learning frameworks, and detect node-level APT attacks in an anomaly detection manner. Experimental results on three real-world APT attack datasets across eight attack scenarios show that the collaborative training model based on multi-view features achieves excellent detection and generalization performance.

\bibliographystyle{unsrt}  
\bibliography{references}

\end{document}